\documentclass[notitlepage,letterpaper]{article}
\usepackage[ansinew]{inputenc} 
\usepackage{amsmath}
\usepackage{amsfonts}
\usepackage{amssymb}
\usepackage{graphicx}
\usepackage{cite}
\usepackage{geometry}      
\usepackage{epstopdf}
\begin{document}
\title{Sound Speeds, Cracking and Stability \\ of \\ Self-Gravitating Anisotropic Compact Objects}
\author{
\textbf{ H. Abreu} \\ 
\textit{Centro de Física Fundamental,} \\ 
\textit{Departamento de Física, Facultad de Ciencias, }\\
\textit{Universidad de Los Andes, Mérida 5101, Venezuela} y \\ 
\textit{Centro Nacional de Cálculo Científico, Universidad de Los Andes,}
\textsc{(CeCalCULA),}  \\
\textit{Corporación Parque Tecnológico de Mérida, Mérida 5101, Venezuela} \\
\textbf{H. Hernández} \\ 
\textit{Laboratorio de Física Teórica,} \\ 
\textit{Departamento de Física, Facultad de Ciencias, }\\
\textit{Universidad de Los Andes, Mérida 5101, Venezuela}\\ and \\ 
\textbf{L. A. Núñez} \\ 
\textit{Centro de Física Fundamental,} \\ 
\textit{Departamento de Física, Facultad de Ciencias, }\\
\textit{Universidad de Los Andes, Mérida 5101, Venezuela} y \\ 
\textit{Centro Nacional de Cálculo Científico, Universidad de Los Andes,}
\textsc{(CeCalCULA),}  \\
\textit{Corporación Parque Tecnológico de Mérida, Mérida 5101, Venezuela}
}
\maketitle

\begin{abstract}
Using the the concept of cracking we explore the influence of density fluctuations and local anisotropy have on the stability of local and  non-local anisotropic matter configurations in general relativity. This concept, conceived to describe the behaviour of a fluid distribution just  after its departure from equilibrium, provides an alternative approach to consider the stability of selfgravitating compact objects. We show that potentially unstable regions within a configuration can be identified as a function of the difference of propagations of sound along tangential and radial directions. In fact, it is found that these regions could occur when, at particular point within the distribution, the tangential speed of sound is greater than the radial one.
\end{abstract}

\section{Introduction}
An increasing amount of theoretical evidence strongly suggests that a variety of very interesting physical phenomena may take place giving rise to local anisotropy, i.e. unequal radial and tangential stresses $P_{r}~\neq~P_{\perp}$ (see  \cite{HerreraSantos1997,MakHarko2003},  and references therein). In the newtonian regime it has been pointed out in the classical paper by J.H. Jeans \cite{Jeans1922}, and in the context of General Relativity, it was early remarked by G. Lema\^{i}tre \cite{Lemaitre1933} that local anisotropy can relax the upper limits imposed on the maximum value of the surface gravitational potential. Since the pioneering work of R. Bowers and E. Liang \cite{BowersLiang1974} its influence in General Relativity has been extensively studied. 

Any model for an anisotropic compact object is worthless if it is unstable against fluctuations of its physical variables and, different degrees of stability/instability will lead to different patterns of evolution in the collapse of self-gravitating 
objects. Therefore, as expected, stability of anisotropic matter configurations in General Relativity has been considered since the beginning of the effort to understand the effects of tangential pressures on a selfgravitating matter configuration \cite{BowersLiang1974}. Very soon, in 1976, W. Hillebrandt and K.O. Steinmetz \cite{HillebrandtSteinmetz1976}, considering the problem of stability  of fully relativistic anisotropic neutron star models, showed (numerically) that there exists a stability criterion similar to the one obtained for isotropic models. Later, Chan, Herrera and Santos\cite{ChanHerreraSantos1993} studied the role played by the local anisotropy in the onset of dynamical instabilities. They found that small anisotropies might drastically change the evolution of the system.  Recently,  an analytical method has been reported to extend the traditional Chandrasekhar's variational formalism \cite{Chandrasekhar1984} to anisotropic spheres \cite{DevGleiser2003}.  

L. Herrera introduces,  in 1992,  the concept of cracking (or overturning) \cite{Herrera1992} which  is a qualitatively different approach to identify potentially unstable anisotropic matter configurations. The idea is that fluid elements, at both sides of the cracking point, are accelerated with respect to each other. It was conceived to describe the behaviour of a fluid distribution just  after its departure from equilibrium.  Later on, Herrera and collaborators \cite{ChanHerreraSantos1993} showed that even small deviations from local isotropy may lead to drastic changes in the evolution of the system as compared with the purely locally isotropic case. More over, they found that  perturbations of density alone, do not take the system out of equilibrium for anisotropic matter configurations. Only perturbations of both, density and local anisotropy induce such departures \cite{DiPriscoEtal1994,DiPriscoHerreraVarela1997}.
This concept refers only to the tendency of the configuration to split (or to compress) at a particular point within the distribution but not to collapse or to expand. The cracking, overturning, expansion or collapse, has to be established from the integration of the full set of Einstein  equations. Nevertheless, it should be clear that the occurrence of these phenomena could drastically alter the subsequent evolution of the system. If within a particular configuration no cracking (or overturning) is to appear, we could identify it as  \textit{potentially} stable (not absolutely stable), because other types of perturbations could lead to its expansions or collapse.

In the present paper we shall explore the influence that fluctuations of density and local anisotropy have on the possible cracking (or overturning) of local and non local anisotropic matter configurations in general relativity. We show that, for particular dependent perturbations, potentially unstable regions within anisotroic matter configurations could occur when the tangential speed of sound, $ {\partial P_{\perp}}/{ \partial \rho} $, is greater than the radial, $ {\partial P_{r}}/{ \partial \rho} $. This can give a more clear physical insight when considering the stability of particular anisotropic configurations when independent perturbations occurs.

This paper is organized as follows. Section \ref{AniMatter} will describe our notation through a brief discussion of local anisotropy matter configurations. The concept of cracking for selfgravitating anisotropic matter configurations and its relation with the sound speeds, is considered in Section \ref{CrackingConf}. The models and modeling strategy are presented in sections \ref{Models} and \ref{Modeling}. Finally some results and conclusions are displayed in Section \ref{Results}.

\section{Anisotropic matter configuration in General Relativity}
\label{AniMatter}
We shall consider a static spherically symmetric anisotropic distribution of matter, described by the Schwarzchild line element  
$\mathrm{d}s^{2} = e^{\lambda(r)} \mathrm{d}t^{2} - e^{\nu(r)} \mathrm{d}r^{2} -r^{2}( \mathrm{d}\theta^{2} + \sin \theta \mathrm{d}\phi^{2})$ and having an energy-momentum tensor represented by $\mathbf{T}_{\nu}^{\mu}~=~\mathrm{diag}~[\rho,-P_{r},-P_{\perp},-P_{\perp}]$, where, $\rho$ is the energy density, $P_{r}$ the radial pressure and $P_{\perp}$ the tangential pressure. For this matter configurations, the general relativistic hydrostatic equilibrium equation can be written \cite{BowersLiang1974} as
\begin{equation}
\frac{\mathrm{d\,}P_{r}}{\mathrm{d}\,r} + \left(  \rho + P_{r}\right)  \left( \frac{m +4\pi r^{3}P_{r}}{r\left(  r-2m \right)  }\right)  - \frac{2}{r} \left(P_{\perp}-P_{r} \right)  = 0. \label{anitov}
\end{equation}
Obviously, in the isotropic case $(P_{\perp}~=~P_{r})$ it becomes the usual Tolman-Oppenheimer-Volkov (TOV) equation, which constrains the internal equilibrium structure of general  relativistic, isotropic, static perfect fluid spheres and it is considered in standard textbooks of gravitation \cite{Weinberg1972,Schutz1985},.

It is clear that the last term in   \ref{anitov}, $\left(  P_{\perp}~-~P_{r}\right)  ~\equiv~\Delta,$ represents a ``force'' due to the local anisotropy. This ``force'' is directed outward when $P_{\perp}~>~P_{r}~\Leftrightarrow~\Delta~>~0$ and inward if $P_{\perp}~<~P_{r}~\Leftrightarrow~\Delta~<~0$. Therefore we should have more massive configurations if $\Delta>0$ and less massive ones if $\Delta<0$. This becomes more evident when the most extreme situations i.e. $P_{\perp}~\neq~0$ and $P_{r}~=~0$ or $P_{\perp}~=~0$ and $P_{r}~\neq~0$ are considered. 

\subsection{Ansatze for an anisotropic equation of state}
If a density profile, $\rho = \rho(r)$ is given, it is possible to integrate \ref{anitov} when the definition of mass and the two other equations of state, 
\begin{equation}
m(r) = 4 \pi \int_{0}^{r} \rho\bar{r}^{2}d\bar{r}, \hspace{0.5cm} P_{r} = P_{r}(\rho)  \hspace{0.5cm} \mathrm{and}  \hspace{0.5cm}
P_{\perp} = P_{\perp} \left( P_{r}  \right) ,
\label{massPrPp}
\end{equation}
are provided. The first equation of state, $P_{r} = P_{r}(\rho)$, corresponds to the standard barotropic equation of state for time-independent systems. In order to close the system, we have also to provide a second equation of state  relating radial and tangential pressures, $P_{\perp} = P_{\perp} \left( P_{r}  \right)$. It has been shown  \cite{RendallSchmidt1991,MarsMartinSenovilla1996},  that there exists a unique global solution to (\ref{anitov}) if: $\rho$ is a continuous positive function; $P_{\perp}(r)$ is a continuous differentiable function; $P_{r}(r)$ is a solution to the equation with starting value $P_{\perp}(0)=P_{r}(0),$ and both pressures are positive at the center (i.e. $P_{\perp}(0)=P_{r}(0)\geq0$), therefore in the whole interior of the body.

Much of the efforts to disentangle the physics of very dense matter is reflected by the various ``radial'' equations of state: $P_{r} = P_{r}(\rho)$ availables (see \cite{HeiselbergPandharipande2000,PageReddy2006} and references therein). In contrast, very little is known for the much less intuitive second equation of state $P_{\perp} = P_{\perp} \left( P_{r}  \right)$. This is the reason why different ansatze are found to introduce anisotropy in matter configurations (see, for instance, references \cite{HerreraSantos1997,MakHarko2003,BowersLiang1974,HeintzmannHillebrandt1975,CosenzaEtal1981,Stewart1982,Rago1991,GokhrooMehra1994,HerreraEtal2001,HerreraMartinOspino2002,DevGleiser2002,HernandezNunez2004,ChaisiMaharaj2005}). The unknown physics in the ``tangential'' equation of state is partially compensated by using heuristic criteria (geometric, simplicity or any other assumption relating radial and tangential pressures). Therefore, most of the exact solutions for the differential equation (\ref{anitov}) found in the literature have been obtained from excessively simplifying heuristic assumptions and, in addition, some of the conditions to become ``physically acceptable fluids'' are not verified.

\subsection{Acceptability conditions for anisotropic matter}
\label{acceptabilitycond}
The interior solution should satisfy some general physical requirements. Some of the ``physical acceptability
conditions'' for anisotropic matter have been stated elsewhere \cite{HerreraSantos1997,MakHarko2003} as
\begin{enumerate}
  \item density, $\rho$, radial pressure, $P_{r}$, and tangential pressure, $P_{\perp}$, should be positive everywhere inside the configuration.
  \item gradients for density and radial pressure should be negative,
  \[
  \frac{\partial \rho }{\partial r} \leq 0 \ ,  \hspace{0.5cm} \mathrm{and} \hspace{0.5cm}
  \frac{\partial P_{r} }{\partial r} \leq 0 \; ;
  \]
  \item inside the static configuration the speed of sound should be less than the speed of light,
  \[ 
  \frac{\partial P_{r} }{\partial \rho} \leq 1 \hspace{0.5cm} \mathrm{and} \hspace{0.5cm}
  \frac{\partial P_{\perp} }{\partial \rho} \leq 1 \; ;
  \]
  \item in addition to the above intuitive physical requirements, the interior solution should satisfy  \cite{KolassisSantosTsoubelis1998} either: 
 \begin{itemize}
  \item the \textit{Strong Energy Condition}:  $\rho+P_{r}+2\,P_{\perp}\geq0$, $\rho+P_{r}\geq0$ and $\rho+P_{\perp }\geq0$  or
  \item the \textit{Dominant Energy Condition}: $\rho   \geq P_{r}$ and $\rho   \geq P_{\perp}$
\end{itemize} 

  \item  junction conditions \cite{BonnorVickers1981}, match the matter configuration to the exterior Schwarzchild solution.  Because of the continuity of the First Fundamental Form, the definition of mass in \ref{massPrPp}, evaluated at the boundary, becomes the total mass, $M = m(a)$, as measured by its external gravitational field. More over, the continuity of the Second Fundamental Form forces the radial pressure to vanish at the boundary, $r =a$, of the sphere $\left. P_{r} \right|_{r=a} = 0$.
\end{enumerate}
Notice that as a consequence of the junction conditions, the radial pressure should vanish at the boundary, but not the tangential one. However, both should be equal at the centre of the matter configuration. Also notice that there is no restriction on the gradient for the tangential pressure.
 
These reasonable physical requirements validate the assumptions made for both equations of state and, in many cases, exclude possible mathematical solutions of the system \ref{anitov} and \ref{massPrPp}. Delgaty and Lake \cite{DelgatyLake1998} considering the isotropic case ($P_{r} = P_{\perp}$), found that,  from 127 published solutions only 16 satisfy the above conditions. In particular, it is worth mentioning that, in order to have a causal theory of matter we have to demand that the sound speed be, at most, the speed of light. This important requirement when obviated, violates physical principles  usually required for a matter physical theory \cite{EllisMaartensMacCallum2007}. 

\section{Instability and cracking of anisotropic compact objects}
\label{CrackingConf}
As we have stressed above, in a series of papers Herrera and collaborators  \cite{Herrera1992, DiPriscoEtal1994, DiPriscoHerreraVarela1997} elaborated the concept of cracking for selfgravitating isotropic and anisotropic matter configurations. It was introduced to describe the behaviour of fluid distributions just after its departure from equilibrium, when total non-vanishing radial forces of different signs appear within the system. 

This section will describe the general framework of the cracking approach to identify potentially stable (and unstable) anisotropic matter configurations. We explicitly use some of the tacit assumptions for modeling cracking (or overturning) within these matter configurations and, finally, we propose a more intuitive criterion based on the difference of sound speeds to estimate the relative magnitude for the density and anisotropy perturbations and to evaluate the stability of bounded distributions.

\subsection{Cracking: the general framework}
Herrera and collaborators state that there is cracking whenever the radial force is directed inward in the inner part of the sphere and reverses its sign beyond some value of the radial coordinate; or, when the force is directed outward in the inner part and changes sign in the outer part, we shall say that there is an overturning. These effects are related to the tidal accelerations of fluid elements 
\cite{DiPriscoEtal1994,Demianski1985}, defined by 
\begin{equation}
a^{\alpha} =  \left[ -R^{\alpha}_{\beta \gamma \mu}u^{\beta} u^{\mu} + h^{\alpha}_{\beta}  \left( \frac{\mathrm{d}u^{\beta}}{\mathrm{d}s} \right)_{;\gamma} - \frac{\mathrm{d}u^{\alpha}}{\mathrm{d}s} \frac{\mathrm{d}u_{\gamma}}{\mathrm{d}s} \right]  h^{\gamma}_{\nu} \; \delta x^{\nu}  \; ,
\label{GeodesicDeviation} 
\end{equation}
where $ \delta x^{\nu}$ is a vector  connecting the two neighbouring particles; $h^{\alpha}_{\beta}$ denotes the projector onto the three-space orthogonal to the four-velocity $u^{\alpha}$  and 
$\mathrm{d}u^{\alpha} / \mathrm{d}s \equiv u^{\mu} u^{\alpha}_{;\mu}$. 
More over, defining
\begin{equation}
 R = \frac{\mathrm{d}P_{r}}{\mathrm{d}r} + 
 \left(\rho + P_{r}\right)\left(\frac{m+4 \pi r^{3} P_{r}}{r\left(r-2m\right)}\right) 
 - \frac{2}{r} \Delta \; ,
 \label{Ranitov}
\end{equation}
it can be shown that \ref{GeodesicDeviation} and  \ref{Ranitov}, evaluated immediately after perturbation, lead to \cite{DiPriscoEtal1994, DiPriscoHerreraVarela1997}
\begin{equation}
R = - \frac{e^{\lambda}(\rho + P_{r})}{e^{\nu/2}r^{2}} \int_{0}^{a} \mathrm{d}\tilde{r} \ e^{\nu/2}\tilde{r}^{2}  \frac{\mathrm{d} \Theta}{\mathrm{d}s} \; ,
\label{Rintegral} 
\end{equation}
where $\Theta$ represents the expansion. Again, here,  
$\mathrm{d}s^{2} = e^{\lambda(r)} \mathrm{d}t^{2} - e^{\nu(r)} \mathrm{d}r^{2} -r^{2}( \mathrm{d}\theta^{2} + \sin \theta \mathrm{d}\phi^{2})$, the static Schwarzchild line element, has been assumed (see reference \cite{DiPriscoEtal1994} for details).  \ref{Ranitov} is just the hydrostatic equilibrium equation \ref{anitov} that vanishes for static (or slowly evolving) configurations. It can be appreciated from (\ref{Rintegral}) that for cracking to occur at some value of $0 \leq r \leq a$, it is necessary that $\mathrm{d} \Theta / \mathrm{d}s$ vanishes somewhere within the configuration. It is also clear the non local nature of this effect; and that small deviations from local isotropy may lead to drastic changes in the evolution of the system as compared with the purely locally isotropic case \cite{DiPriscoHerreraVarela1997}. 

\subsection{Cracking revisited}
Following \cite{DiPriscoHerreraVarela1997}, we assume that the system having some pressure and density distributions satisfying $R =0$, is perturbed from its hydrostatic equilibrium. Thus, fluctuations in density and anisotropy induce total radial forces ($R \neq 0$) which, depending on their spatial distribution, may lead to the \textit{cracking}; i.e radial force directed inward, ($R > 0$),  or, \textit{overturning}, directed outward, ($R < 0$) of the source. Therefore, we will be looking for a change of the  sign of $R$, beyond some value of the radial coordinate. We will exclusively consider perturbations on both, density and local anisotropy, under which the system will be dynamically unstable.  In other words,  $\delta \rho$ and  $\delta \Delta$ are going to be considered as independent perturbations; but fluctuations in mass and radial distribution pressure depend on density perturbations, i.e.
\begin{equation}
\label{EmeP}
 \rho + \delta \rho \Rightarrow 
\left\{\begin{array}{l}
 P_{r}(\rho +\delta \rho,r)   \approx P_{r}(\rho,r) + \delta P_{r}  \approx P_{r}(\rho,r) + \frac{\partial P_{r} }{ \partial  \rho} \delta \rho  \; ,\\
       \\
 m(\rho +\delta \rho,r) = 4 \pi \int_{0}^{r} (\rho + \delta \rho) \bar{r}^{2}d\bar{r}  \approx 
 m(\rho,r) + \frac{4 \pi}{3}r^{3} \delta \rho \; .
\end{array}
\right.
\end{equation}
Now on, expanding \ref{Ranitov}  we have, formally,
\begin{equation}
\label{Rexpanded}
R \approx R_{0}(\rho,P_{r},m,\Delta,r) +\underbrace{\frac{\partial R}{ \partial \rho} \delta \rho 
+\frac{\partial R}{ \partial P_{r}} \delta P_{r} 
+\frac{\partial R}{ \partial m} \delta m
+\frac{\partial R}{ \partial \Delta} \delta \Delta}_{\tilde{R}}
\end{equation}
and by using \ref{EmeP} it can be shown that
\begin{equation} 
\tilde{R} =  \delta \rho \left[ \left(2 \frac{\partial R}{ \partial \rho} + \frac{4 \pi}{3}r^{3} \frac{\partial R}{ \partial m} \right) - \frac{2}{r} \frac{\delta \Delta}{\delta \rho} \right] \; ,
\label{fractura}
\end{equation}
where
\begin{equation}
\label{DifRrhoDifRm}
\frac{\partial R }{ \partial  \rho} = \frac {m + 4 \pi \,P_{r} r^{3}}{r ( r-2m ) } \geq 0  \hspace{0.3cm} \mathrm{and} \hspace{0.3cm} 
 \frac{\partial R}{ \partial m} = \frac{ ( \rho+P_{r} )  \left( 1 + 8\pi \,P_{r} r^{2} \right) }{ ( r-2\,m )^{2}} \geq 0 \; .
\end{equation}

It is immediatelly seen that, in order to have $\tilde{R} = 0$ and consequently  a change in its sign:  
\begin{itemize}
  \item both, the anisotropy and the density, have to be perturbed;
  \item both, anisotropy and density perturbations, have to have the same sign, i.e. $\delta \Delta / \delta \rho > 0$. 
\end{itemize}
In order words, potentially stable configurations should have $\delta \Delta / \delta \rho \leq 0$ everywhere because $\tilde{R}$ never changes its sign \cite{AbreuHernandezNunez2007}. 

\subsection{Cracking and sound speeds}
When arbitrary and independent density and anisotropy perturbations are considered (as in all previous works concerning cracking \cite{Herrera1992,DiPriscoEtal1994,DiPriscoHerreraVarela1997,AbreuHernandezNunez2007}) there is little physical criteria to establish the size (absolute and/or relative) of the perturbation, i.e. how small (or big) the perturbations should be. Different orders of magnitude (and relative size  $\delta \Delta/ \delta \rho$) of perturbations could produce a cracking but we could be describing an unphysical scenario. Additionally, all these previous works only consider constant perturbations. It is possible that variable perturbations could be more efficient inducing cracking within a particular matter configuration. Again, there is no criteria in establishing the functionality of the perturbation throughout the matter distribution. 

We are going to consider a particular type of dependent perturbation whose relative order of magnitude could be bounded by the behavior of some physical variables and could be checked by physical intuition. Obviously, general perturbations should be independent because they emerge from non related physical phenomena. But we are looking for some physical variables whose behavior could be checked in order to identify potential cracking. 

It is easy to convince oneself that
\begin{equation}
 \frac{\delta \Delta}{\delta \rho} \sim \frac{\delta  \left(P_{\perp}-P_{r} \right)}{ \delta \rho } \sim
 \frac{\delta P_{\perp}}{ \delta \rho} - \frac{\delta P_{r}}{ \delta \rho} \sim v^{2}_{s \perp} - v^{2}_{s r}  \; \; ,
\label{SoundRelation}
\end{equation} 
where $v^{2}_{s  r}$  and $v^{2}_{s \perp}$ represent the radial and tangential sound speeds, respectively. 

This will be the key concept, we will use in revisiting Herrera's approach to identify potentially unstable anisotropic matter configurations based on the concept of cracking. Now, by considering the sound speeds and evaluating \ref{SoundRelation},  we could not only have a more precise idea of the relative order of magnitude of the perturbations ($\delta \Delta$ and  $\delta \rho$)  but also what are the regions more likely to be potentially unstable within a matter configuration. 

It is clear that, because $0 \leq  v^{2}_{s r} \leq 1$ and $0 \leq v^{2}_{s \perp} \leq 1$, we have 
$| v^{2}_{s \perp} - v^{2}_{s r}|  \leq 1$. Thus,
\begin{equation}
-1 \leq v^{2}_{s \perp} - v^{2}_{s r}  \leq 1  \Rightarrow 
\left\{
\begin{array}{c l}
  -1 \leq v^{2}_{s \perp} - v^{2}_{s r} \leq 0   &  \mathrm{ Potentially \; stable} \; , \\
     &    \\      
0 <  v^{2}_{s \perp} - v^{2}_{s r}  \leq 1   &    \mathrm{ Potentially \; unstable} \; .
\end{array}
\right.     
\label{VelocityStability1}
\end{equation}
Therefore, we can now evaluate potentially unstable regions within anisotropic models  based on the difference of the propagation of sound within the matter configuration. Those regions where $ v^{2}_{s r}  > v^{2}_{s \perp}$ will be potentially unstable. On the other hand, if $ v^{2}_{s r}  \leq v^{2}_{s \perp}$ everywhere within a matter distribution, no cracking will occur. It is worth mentioning, concerning this criterion, one of the extreme matter configurations mentioned above ($P_{\perp}~\neq~0$ and $P_{r}~=~0$) is allways potentially stable for cracking; and the other one ($P_{\perp}~=~0$ and $P_{r}~\neq~0$) becomes potentially unstable.

More over, for physically reasonable models, the magnitude of perturbations in anisotropy should always be smaller than those in density,  i.e. $|v^{2}_{s \perp}~-~v^{2}_{s r}|~\leq~1\Rightarrow |\delta \Delta | \leq |\delta \rho|$. When $\delta \Delta / \delta \rho > 0$, these perturbations lead to potentially unstable models. 

Next section will be devoted to explore the effectiveness of this criterion on the stability of bounded matter configurations, having different equations of state.  Again, we recall that the concept of cracking refers only to the tendency of the configuration to split and its occurrence has to be established from the integration of the full set of Einstein equations. In addition, it is clear that there could also be some perturbations that do not induce cracking but could cause instabilities that lead the configuration to collapse or to expand. 

\section{Perturbations and cracking for anisotropic configurations}
\label{Models}
In order to illustrate the workability of the above criterion \ref{VelocityStability1}, we shall work out several density profiles for models satisfying the physical acceptable conditions. Thus, in addition to the positivity of density and pressures profiles, their  gradients and the fulfillment of the energy conditions (strong or dominant), we shall pay special and particular attention to the conditions bounding sound speeds (radial and tangential) within the matter configuration. 

The idea will be to provide the density profile; then to obtain the radial pressure, $P_{r}(r)$ from a ``radial'' equation of state $P_{r} = P_{r}(\rho(r))$ and next to solve the tangential pressure $P_{\perp}(r)$ from the anisotripic TOV  \ref{anitov}.  That is.
\begin{equation}
\label{Density2PressureProfile}
\rho(r) \rightarrow P_{r} = P_{r}(\rho(r))  \rightarrow P_{\perp} =  P_{r}  + \frac{r}{2} \frac{\mathrm{d}P{r}}{\mathrm{d}r} +
\frac{ \left(\rho + P_{r}\right)}{2}\left(\frac{m+4 \pi r^{3} P_{r}}{ \left(r-2m\right)}\right) \; .
\end{equation}
Then, the radial and tangential sound speeds are calculated and their difference $|v^{2}_{s \perp}~-~v^{2}_{s r}|$ is evaluated. Next, by using \ref{VelocityStability1} the pontential stability or unstability is established. This will be confirmed by a change in the sign of $\tilde{R}$ described by \ref{fractura} and \ref{DifRrhoDifRm}. 

To illustrate the above criterion \ref{VelocityStability1} we shall analyze four cases concerning qualitatively  different density profiles. We have selected two local (one singular and one non singular)  and two non-local conformally flat anisotropic solutions.
By \textit{local models} we mean the standard way to express an equation of state where the energy density and radial pressure are related at a particular point within the configuration, i.e. $P = P(\rho(r))$. On the other hand, by \textit{non local models} we will understand those where the radial pressure $P_{r}(r)$ is not only a function of the energy density, $\rho(r),$ at that point; but also its functional throughout the rest of the configuration.  Any change in the radial pressure takes into account the effects of the variations of the energy density within the entire volume \cite{HernandezNunezPercoco1998,HernandezNunez2004,MunozNunez2006}.  It has been shown that in the static limit, this particular radial equation of state can be written as
\begin{equation}
P_{r}(r)=\rho(r)-\frac{2}{r^{3}}\int_{0}^{r}\bar{r}^{2}\rho(\bar{r})\ \mathrm{d}\bar{r} \; . 
\label{globalst}
\end{equation}
It is clear that in equation (\ref{globalst}) a collective behavior of the physical variables $\rho(r)$ and $P_{r}(r)$ is present.

\subsection{Anisotropic Tolman VI model}
This model was introduced by Cosenza, Herrera, Esculpi and Witten \cite{CosenzaEtal1981} starting from the singular Tolman VI density profile\cite{Tolman1939}. The original isotropic Tolman VI solution is not deprived of a physical meaning. It resembles a highly relativistic Fermi Gas with the corresponding adiabatic exponent of 4/3. By using a heuristic method these authors determine other physical variables representing an anisotropic static matter configuration; i.e.
\begin{equation} \label{VarPhyTVI}
\rho=\frac{K}{r^{2}}, \quad \Rightarrow P_{r}=\frac{3}{8\pi r^{2}}\left( \frac{1-\sqrt{\frac{r}{a}}}{7-3\sqrt{\frac{r}{a}}} \right), \quad \Rightarrow P_{\perp}=\frac{3}{224 \pi r^{2}} \left( \frac{21-25\sqrt{\frac{r}{a}}}{7-3\sqrt{\frac{r}{a}}} \right)
\end{equation}
where, the junction conditions force the adjustment of the parameter, to $K= {3}/{56 \pi}$,  and the radius is given by $a= {81}/{49}$. 

Sound speeds can be determined from  \ref{VarPhyTVI},  and can be written as 
\begin{equation} \label{VelSonRadTVI}
  v^{2}_{s r}=\frac{7\left(7+3\frac{r}{a}-9\sqrt{\frac{r}{a}}\right)}{\left(7-3\sqrt{\frac{r}{a}}\right)^{2}} \quad \textrm{and} \quad
v^{2}_{s \perp }=\frac{3\left(49+25\frac{r}{a}-70\sqrt{\frac{r}{a}}\right)}{4\left(7-3\sqrt{\frac{r}{a}}\right)^{2}} \; .
\end{equation}

By using \ref{VarPhyTVI} the particular expresion for \ref{fractura} can be obtained for this model as
\begin{equation}\label{FractTolmanVI}   
\tilde{R}_{TolmanVI}=\frac{2\delta\rho}{r}\left[\frac{\left(588+180\frac{r}{a}-672\sqrt{\frac{r}{a}}\right)v^{2}_{s, \ r} +539+195\frac{r}{a}-658\sqrt{\frac{r}{a}}}{16\left(7-3\sqrt{\frac{r}{a}}\right)^{2}} -\frac{\delta \Delta}{\delta \rho}\right] \; .
\end{equation}

It is worth mentioning that this model does not fulfill  all the acceptability conditions stated in \ref{acceptabilitycond}.   It is singular at the center and near the boundary surface  ($r/a \simeq 0.706$) the tangential pressure becomes negative. Despite this unphysical situation, this model is presented because, as it will later become clear in Section \ref{Modeling}, the difference in sound speed is constant through the whole configuration, which represents the above mentioned constant perturbation relation considered in previous works  \cite{Herrera1992,DiPriscoEtal1994,DiPriscoHerreraVarela1997,AbreuHernandezNunez2007}.

\subsection{Non local Stewart Model 1}
This model emerges from a density profile proposed by B. W. Stewart \cite{Stewart1982}, to describe anisotropic conformally flat static bounded configurations; which was also, recently, considered for non local anisotropic  matter distributions \cite{HernandezNunez2004}. Starting from this density profile we can find $P_{r}(r)$ and $P_{\perp}(r)$ as
 
\begin{eqnarray} 
\rho=\frac{1}{8\pi r^2} \frac{(e^{2Kr}-1)(e^{4Kr}+8Kre^{2Kr}-1)}{(e^{2Kr}+1)^{3}} \label{Stewart1Dens} \\
 \Downarrow \nonumber \\ 
P_{r}=\frac{1}{8\pi r^2}\frac{(1-e^{2Kr})(e^{4Kr}-8Kre^{2Kr}-1)}{(e^{2Kr}+1)^{3}}  \label{Stewart1PresR} \\
     \Downarrow \nonumber \\
     P_{\perp}=\frac{2K^{2}e^{4Kr}}{\pi [1+e^{2Kr}]^{4}} \label{Stewart1PresT} \; .
\end{eqnarray}
 
Again, the parameter $K$ has to be obtained from the junction conditions; which means that $K$ has to satisfy a trascendental equation 
\begin{equation} \label{KStewart}
e^{4Ka}-8Kae^{2Ka}-1=0 \quad \Rightarrow \quad K=\frac{1}{2a}\ln\left[\frac{1+\left(\frac{2M}{a}\right)^{\frac{1}{2}}}{1-\left(\frac{2M}{a}\right)^{\frac{1}{2}}}\right] 
  \; .
\end{equation}
From \ref{Stewart1Dens}, \ref{Stewart1PresR} and \ref{Stewart1PresT} the corresponding sound speeds can be found
 \begin{equation}\label{VsRStewart}
        v^{2}_{s r}=\frac{8Kre^{2Kr}\left[\left(e^{4Kr}-1\right)+Kr\left[\left(e^{2Kr}-2\right)^{2}-3\right]\right]-\left(e^{4Kr}-1\right)^{2}}{\left(e^{4Kr}-1\right)^{2}+8K^{2}r^{2}e^{2Kr}\left[\left(e^{2Kr}-2\right)^{2}-3\right]}
    \end{equation} and
 \begin{equation} \label{VsTStewart}
        v^{2}_{s \perp}=\frac{32K^{3}r^{3}e^{4Kr}\left(e^{2Kr}-1\right)\left(e^{2Kr}+1\right)^{-1}}{\left(e^{4Kr}-1\right)^{2}+8K^{2}r^{2}e^{2Kr}\left[\left(e^{2Kr}-2\right)^{2}-3\right]} \; .
    \end{equation}
        
Now, equation  \ref{fractura} can also be obtained for this model as
\begin{equation}\label{FractNLStewart1}
\tilde{R}_{NLStewart1} =\frac{2\delta\rho}{r}\left[ \frac{\left[\left(e^{2Kr}+1\right)\left(5+v^{2}_{s r}\right)+4Kr\left(e^{2Kr}-1\right)\right]Kr}{6 \left(e^{2Kr}+1\right)^{2}\left(e^{2Kr}-1\right)^{-1}}   -\frac{\delta\Delta}{\delta\rho}  \right] \; .
\end{equation}

\subsection{Non local Stewart Model 2}
This is a second density profile proposed by B. W. Stewart \cite{Stewart1982} which was recently proved to be non local \cite{MunozNunez2006} :
 
\begin{eqnarray} 
\rho =  \frac{1}{8\pi r^{2}} \left[ 1 - \frac{\sin(2K r)}{K r} + \frac{\sin^{2}(K r)}{K^{2}r^{2}}\right] \label{Stewart2Dens} \\
 \Downarrow \nonumber \\ 
P_{r} =  -\frac{1}{8\pi r^{2}}\left[1+\frac{\sin(2K r)}{K r}-3\frac{\sin^{2}(K r)}{K^{2}r^{2}} \right]   \label{Stewart2PresR}\\
     \Downarrow \nonumber \\
     P_{\perp} =  \frac{1}{8\pi r^{2}}\left[1-\frac{\sin^{2}(K r)}{K^{2} r^{2}} \right] \; . \label{Stewart2PresT}
\end{eqnarray}
 
As in the previous models, junction conditions ($M=m(a)$ and $P_{r}(a)=0$) determine the coupling constant $K$. In this case it has to satisfy also a trascendental equation
\begin{equation} \label{AcopleStewart2}
\frac{\sin Ka}{Ka}=\left(1-\frac{2M}{a}\right)^{1/2} \quad \mathrm{and} \quad 
\cos Ka =\frac{1 - \frac{3M}{a}}{\sqrt{1 - \frac{2M}{a}}} \; .
\end{equation} Thus,
\begin{equation}
K = \frac{\sqrt{\frac{M}{a}\left( 4 - \frac{9M}{a} \right)}}{a - 2M} \, .
\end{equation}

From \ref{Stewart2Dens}, \ref{Stewart2PresR} and \ref{Stewart2PresT} the corresponding sound speeds can be found
 \begin{equation} \label{VsRStewartII}  
        v^{2}_{s r}=\frac{\sin^{2}(Kr)\left(3-K^{2}r^{2}\right)-\frac{3}{2}\sin(2Kr)Kr}{\left[\cos(Kr)Kr - \sin(Kr)\right]^{2}} 
        \quad \textrm{and} \quad v^{2}_{s \perp}=\frac{\frac{Kr}{2}\left[Kr+\frac{\sin(2Kr)}{2}\right]-\sin^{2}(Kr)}{\left[\cos(Kr)Kr - \sin(Kr)\right]^{2}} \; .
    \end{equation} 
    
Now, equation  \ref{fractura} can also be obtained for this model as
\begin{equation}\label{FractNLStewart2}
\tilde{R}_{NLStewart2} =\frac{2\delta \rho}{r}\left[\frac{2\sin(2Kr)Kr-9\sin^{2}(Kr)\left(v^{2}_{s r}+1\right)}{12\sin^{4}(Kr)\left[\sin(2Kr)Kr-2\sin^{2}(Kr)\right]^{-1}}-\frac{\delta \Delta}{\delta \rho}\right] \; .
\end{equation}

\subsection{Florides-Stewart-Gokhroo \& Mehra Model}
This density profile is due originally to P.S. Florides \cite{Florides1974}, but also corresponds to different solutions, considered by Stewart \cite{Stewart1982} and, more recently, by M. K. Gokhroo and A. L. Mehra \cite{GokhrooMehra1994}. The Florides-Stewart-Gokhroo-Mehra solution represents densities and pressures which, under particular circumstances \cite{Martinez1996}, give rise to an equation of state similar to the Bethe-B\"{o}rner-Sato newtonian equation of state for nuclear matter
\cite{Demianski1985,ShapiroTeukolsky1983,BetheBornerSato1970}.
 
\begin{eqnarray} 
\rho = \rho=\rho_{c}\left(1-\frac{Kr^{2}}{a^{2}} \right)  \label{GokhrooDens} \\
 \Downarrow \nonumber \\ 
P_{r} = \frac{\rho_{c}}{j}\left(1-\frac{2\mu r^{2}}{a^{2}}\left[\frac{5-\frac{3K r^{2}}{a^{2}}}{5-3K}\right] \right)\left(1-\frac{r^{2}}{a^{2}}\right)^{n},   \label{GokhrooPresR}\\
     \Downarrow \nonumber \\
     P_{\perp} = P_{r}+\frac{\rho_{c}}{j} \left[ \frac{3\mu K}{5-3K} \eta^{4} \left(1- \eta^{2}  \right)^{n} +  \eta^{2} \frac{e^{\lambda}}{2} \left [ \frac{15\mu e^{-2\lambda}}{j\left(5-3K\right)}  \left(1 - \eta^{2} \right)^{2n}\right.\right. 
     \nonumber \\
  \qquad \qquad  \left. \left.-2\pi e^{-2\lambda} \left(1- \eta^{2} \right)^{n-1} + \frac{5\mu j}{5-3K}\left(1-\frac{3}{5} K\eta^{2} \right)\left(1-K\eta^{2} \right) \right]\right] \; ,  \label{GokhrooPresT}
\end{eqnarray}
   
with 
\[
\mu=\frac{M}{a}, \quad  e^{-\lambda}=1-\frac{2\mu\eta^{2}\left(5-3K\eta^{2}\right)}{5-3K}  
 \quad \textrm{and} \quad \eta=\frac{r}{a} .
\] and $\rho_{c}$ the density at the center of the matter configuration.

From \ref{GokhrooDens}, \ref{GokhrooPresR} and \ref{GokhrooPresT} the corresponding radial sound speed can be found as 
    \begin{equation} \label{VsRGokhroo}
        v^{2}_{s r}=\frac{2\mu\left(5-6K \eta^{2} \right)\left[1-\left( 1+n \right)\eta^{2} \right]+n\left[5-3K\left(1+2\mu \eta^{4} \right) \right]}{Kj\left( 5-3K \right)\left(1- \eta^{2} \right)^{1-n}}
    \end{equation} and
        \begin{equation} \label{VsT-VsR_Gokhroo}
 v^{2}_{s \perp}= v^{2}_{s r} -\frac{1}{2jK\eta}\left[ \frac{6\mu K \eta^{3}\left[2-\eta^{2}(2+n)\right]}{(5-3K)\left(1-\eta^{2}\right)^{1-n}}+\frac{\eta }{e^{-\lambda}}\left[\phi\left(1-\frac{\eta \xi}{2e^{-\lambda}}\right) +\frac{\eta}{2}\Xi \right] \right] \; ,
    \end{equation} 
    where
\begin{equation}  
        \Xi=\frac{30\mu e^{-\lambda}\left(1-\eta^{2}\right)^{2n}}{j(5-3K)}\left(\xi -\frac{2n\eta e^{-\lambda}}{1-\eta^{2}}\right)
           -\frac{4\mu j K \eta \left(4-3K\eta^{2}\right)}{5-3K} -4\pi e^{-\lambda}\left[\xi -\frac{\eta e^{-\lambda}}{1-\eta^{2}}\right]\left(1-\eta^{2}\right)^{n-1}
 \end{equation}
     \begin{equation} 
        \phi=\frac{1}{5-3K}\left[\left(1-\eta^{2}\right)^{2n}e^{-2\lambda}\left[\frac{15 \mu}{j}-\frac{2\pi\left(5-3K\right)}{ \left(1-\eta^{2}\right)^{n+1}}\right] +  \frac{ \mu j \left(5-3K\eta^{2}\right)}{\left(1-K\eta^{2}\right)^{-1}} \right] 
    \end{equation} and
    \begin{equation}
        \xi=-\frac{4\mu\eta\left(5-6K\eta^{2}\right)}{5-3K}.
    \end{equation}

For this example equation  \ref{fractura} can also be obtained as
\begin{eqnarray}\label{FractGokhroo}
\tilde{R}_{FSGM}  =\frac{2\delta \rho}{r}\left[\frac{2\mu v^{2}_{s r}}{\left(5-3K\right)je^{-2\lambda}a\eta}\Bigg[\eta^{2}\left[15 e^{-\lambda}\left(1-\eta^{2}\right)^{n}-j\left(5-6K\eta^{2}\right)\right] \right. \nonumber \\ \qquad \qquad
     - \left. \left.\frac{2\mu \left(5-3K\eta^{2} \right)\left[15e^{-\lambda}\left(1-\eta^{2}\right)^{n}+j\left(10-9K\eta^{2}\right)   \right]}{5-3K}                 \right]  \right. \\ \qquad 
     -\frac{6\mu \eta^{4}}{\left(5-3K\right)^{2}j^{2}} \Bigg[ 
        j^{2}\left(9K^{2}\eta^{4}+25\right)+30j\left[e^{-\lambda}\left(1-\eta^{2}\right)K-j\right]\eta^{2} \nonumber \\
         \qquad \qquad \qquad \qquad \qquad \qquad
        -75e^{-2\lambda}\left(1-\eta^{2}\right)^{2n} \Bigg]-\frac{\delta \Delta}{\delta \rho}\Bigg] \; . \nonumber 
\end{eqnarray}

\begin{table}
\begin{center}%
\begin{tabular}{cccccc}
\textbf{Density Profile} & $M/a$ & $M(M_{\odot})$ & $z_{a}$ & $\rho_{a}$ $\times$ $10^{14}\,(gr/cm^{3})$ & $\rho_{c}$ $\,\times10^{15}$ $(gr/cm^{3})$  \\
\textbf{\textit{TolmanVI }} 		& 0.21 & 1.42 & 0.31 & 2.30 & NA\\ 
\textbf{\textit{NL Stewart 1}} 	& 0.32 & 2.15 & 0.65 & 6.80 & 1.91 \\ 
\textbf{\textit{NL Stewart 2}} 	& 0.39 & 2.68 & 1.19 & 8.49 & 2.14 \\
\textbf{\textit{Gokhroo \& Mehra}} & 0.26 & 1.76 & 0.44 & 0.00 & 2.09 \\
\end{tabular}
\end{center}
\caption{All  parameters have been chosen to represent a possible compact object with $a=10$ Km. and the corresponding mass function satisfying the physical acceptability and energy conditions}
\label{ModParameter}
\end{table}

\section{The modeling performed}
\label{Modeling}
As it can be appreciated from the parameters displayed in table \ref{ModParameter}, all the models considered have radius ($a=10$ Km.) and total masses, $M$ (in terms of solar mass $M_{\odot}$) that correspond to typical values for expected astrophysical compact objects. The boundary redshifts $z_{a}$, surface and central densities, $\rho_{a}$ and $\rho_{c}$ that emerge from our selection, also fit the typical values for these objects.
\begin{figure}
  \begin{center}
    \begin{tabular}{cc}
      \resizebox{59mm}{!}{\includegraphics[angle=270]{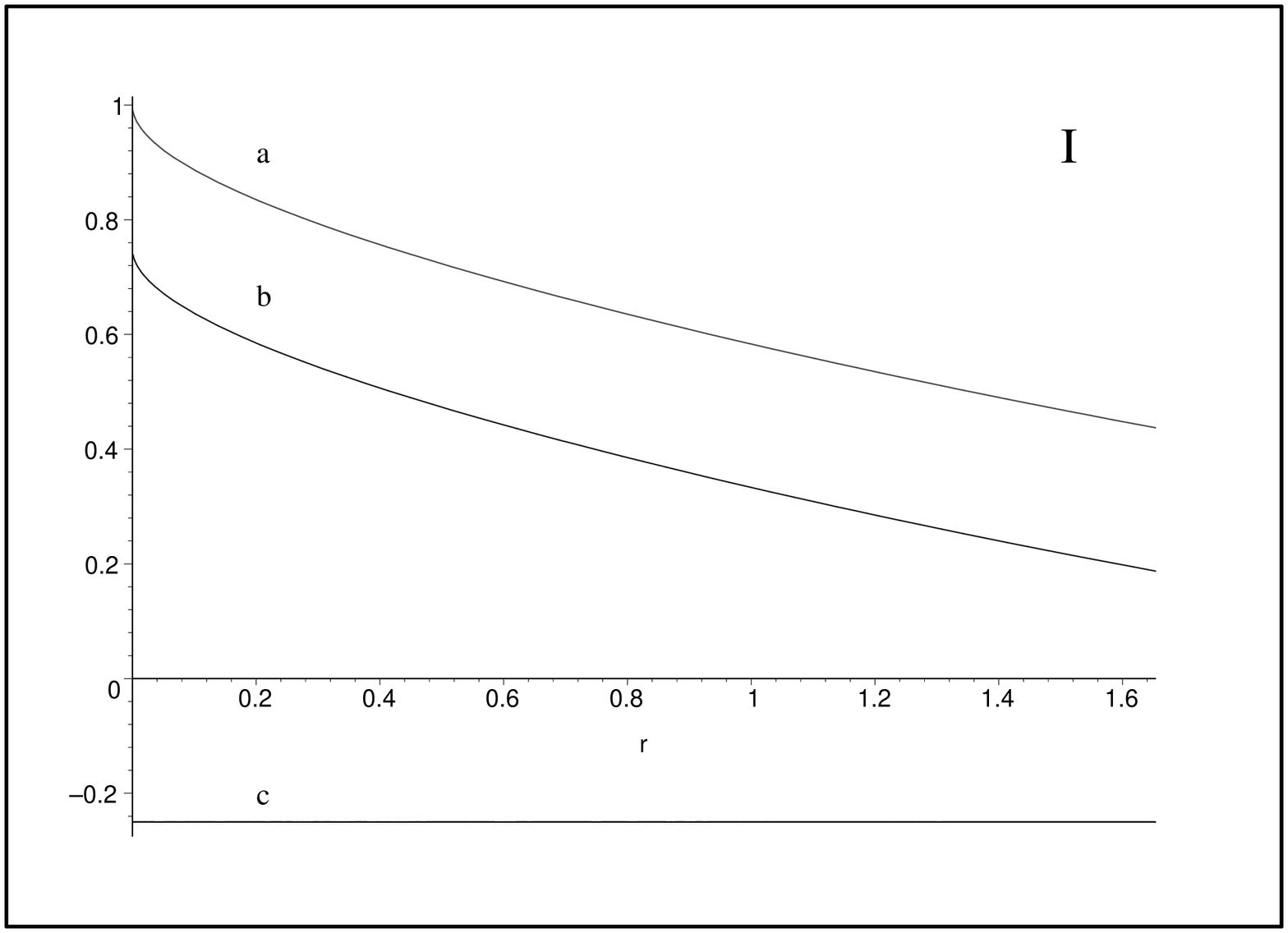}} &
      \resizebox{60mm}{!}{\includegraphics[angle=270]{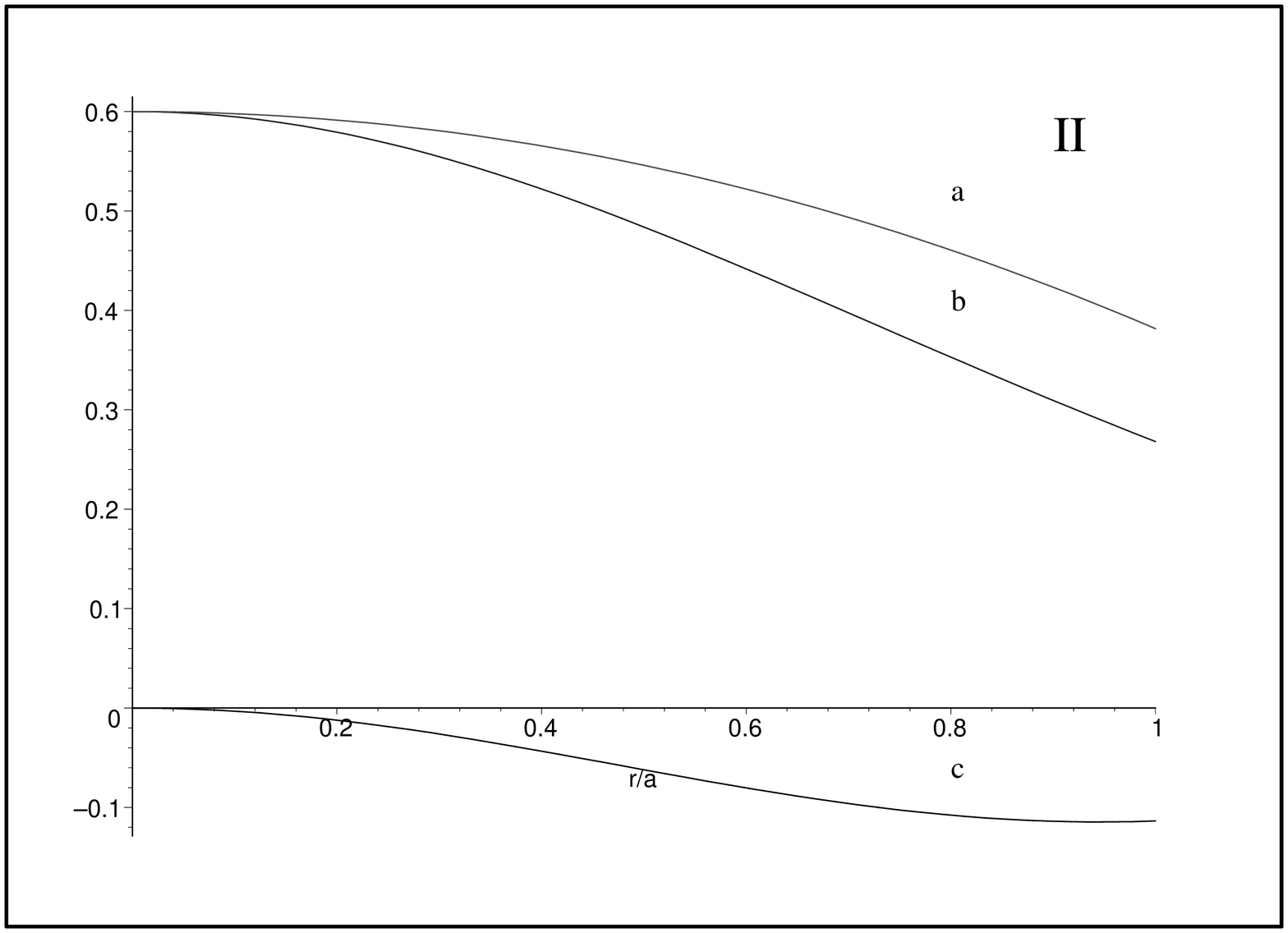}} \\
      \resizebox{59mm}{!}{\includegraphics[angle=270]{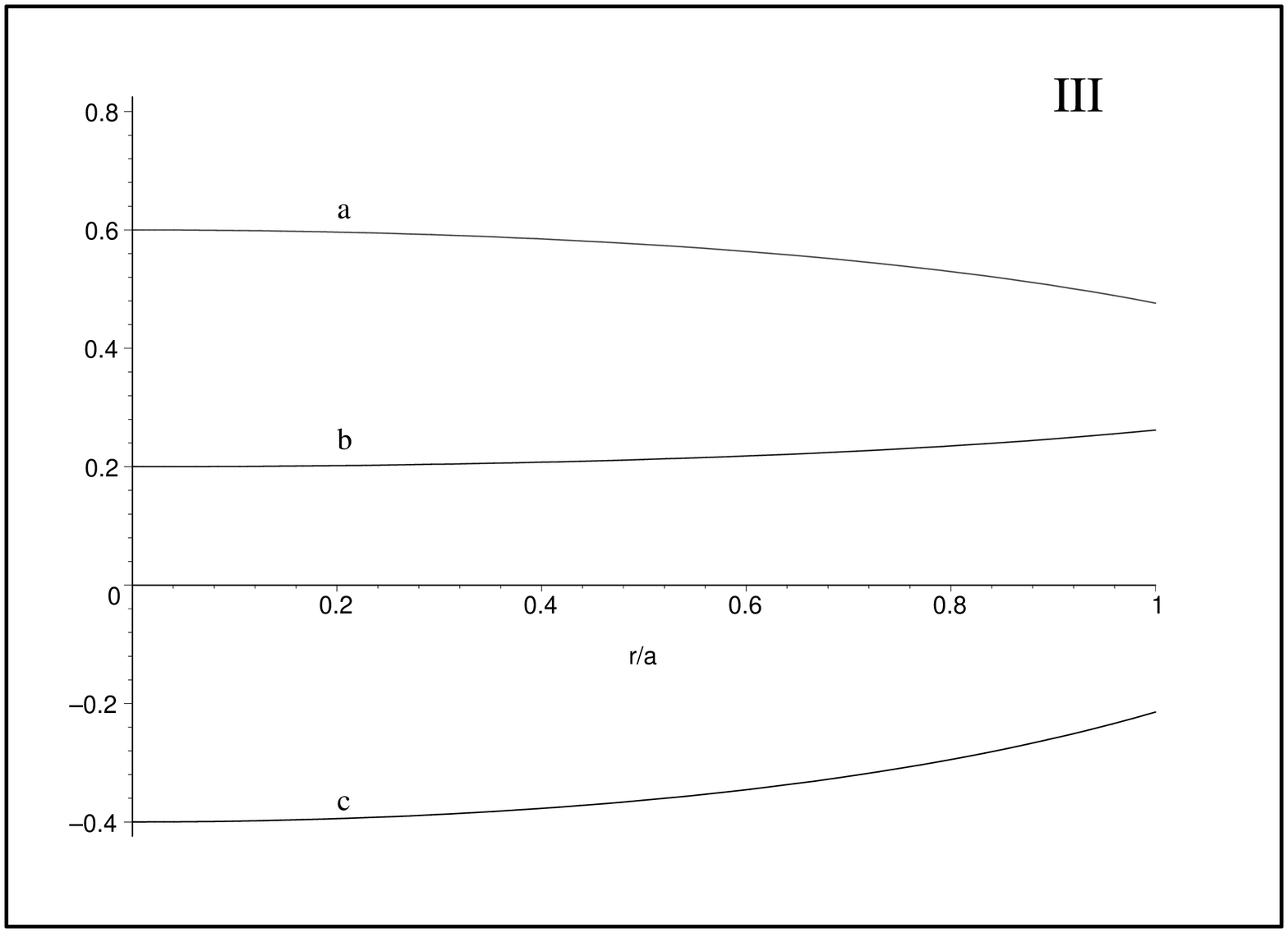}} &
      \resizebox{60mm}{!}{\includegraphics[angle=270]{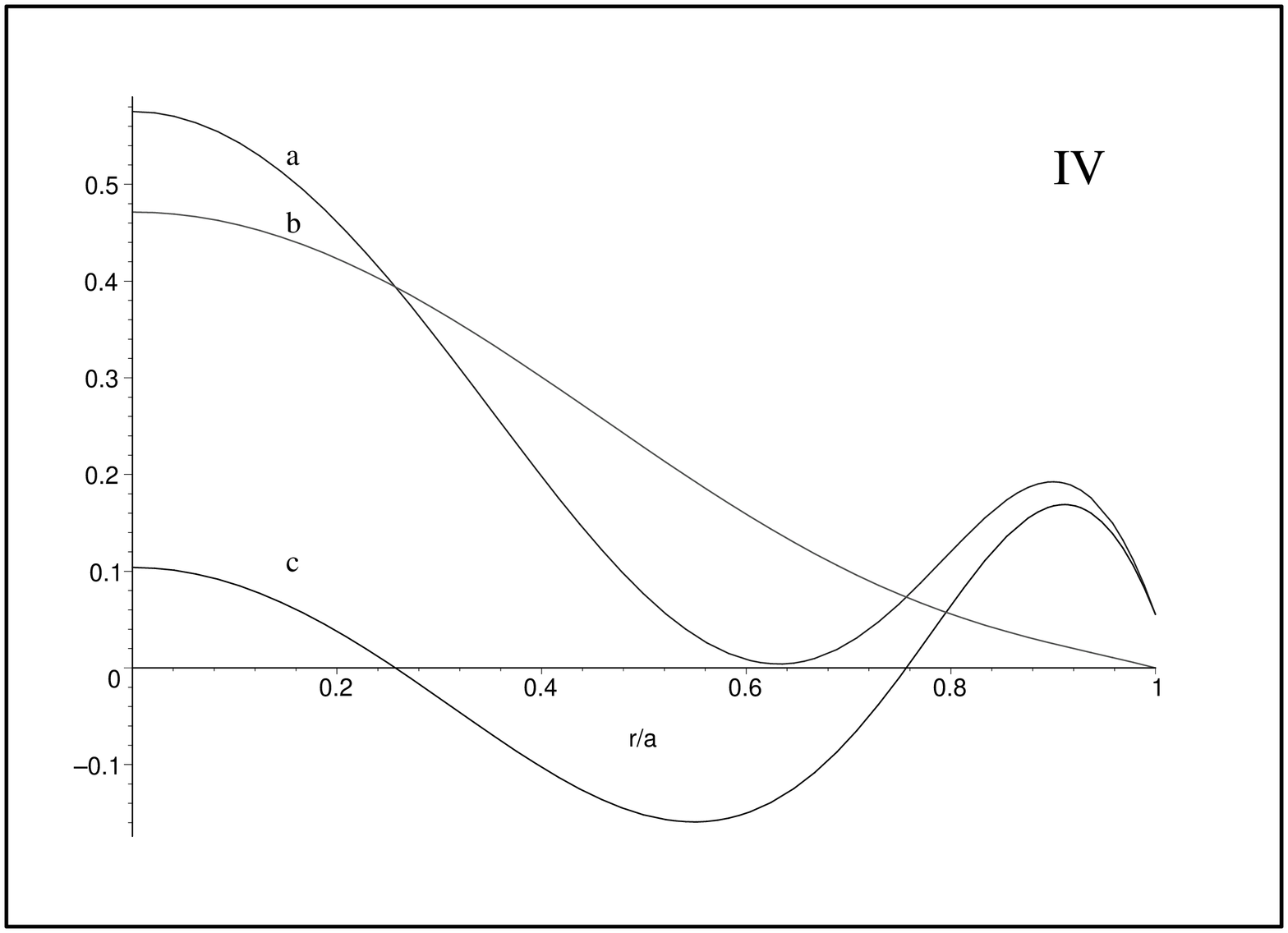}} \\
    \end{tabular}
    \hspace{2pc}
    \begin{minipage}[b]{35pc}
        \caption{Variations of the radial and tangential sound speeds for anisotropic configurations. Plates $I$, $II$, $III$ y $IV$ represent Tolman VI, NL Stewart 1, NL Stewart 2 and Gokhroo \& Mehra, respectively. Curves $a$, $b$ and $c$  correspond to $v^{2}_{s r}$, $v^{2}_{s \ \bot}$ y $v^{2}_{s \ \bot}-v^{2}_{s r}$, respectively. }
        \label{VelSon}
    \end{minipage}
  \end{center}
\end{figure}

Profiles for the radial, $v^{2}_{s r}$, and tangential, $v^{2}_{s \perp}$,  sound speeds, as well as its difference, $v^{2}_{s \perp} - v^{2}_{s r} $, are displayed in Figure \ref{VelSon}. The perturbation relation, $ {\delta \Delta}/{\delta \rho} \equiv v^{2}_{s \perp} - v^{2}_{s r}$ fulfills the physical restriction  $-1 \leq {\delta \Delta}/{\delta \rho}   \leq 1 $ for all models considered. Notice that ${\delta \Delta}/{\delta \rho}$, is constant within the matter distribution for the Tolman VI anisotropic model (plate $I$ in Figure \ref{VelSon}). This type of constant perturbation relations were standard for modeling cracking in previous works \cite{Herrera1992, DiPriscoEtal1994, DiPriscoHerreraVarela1997,AbreuHernandezNunez2007}. Because ${\delta \Delta}/{\delta \rho} < 0$, the sound speed stability criterion, \ref{VelocityStability1}, states that in the Tolman VI anisotropic model, no cracking will occur. Non local Stewart models are sketched in plates $II$ and $III$, respectively. For these two models we could implement variable perturbation relations, ${\delta \Delta}/{\delta \rho}$, through the matter configuration; because  $-1 \leq {\delta \Delta}/{\delta \rho}   \leq 0$, no cracking will occur in these models either. Finally, the most interesting scenario emerges from the Florides-Stewart-Gokhroo-Mehra model \cite{GokhrooMehra1994} with $j =7$, $K=1$ and $n =2$, shown in plate $IV$. As it is evident from this plate, the perturbation relation, $ {\delta \Delta}/{\delta \rho}$ not only has a variable profile, but it also changes its sign, alternating potentially stable and unstable regions within the distribution. In fact, this model presents two potentially unstable regions: $ 0 \lesssim  \eta = r/a \lesssim  0.2570$ and $0.7565 \lesssim  \eta = r/a \lesssim 1$ where $ {\delta \Delta}/{\delta \rho} > 0$.

\begin{figure}
  \begin{center}
    \begin{tabular}{cc}
      \resizebox{59mm}{!}{\includegraphics[angle=270]{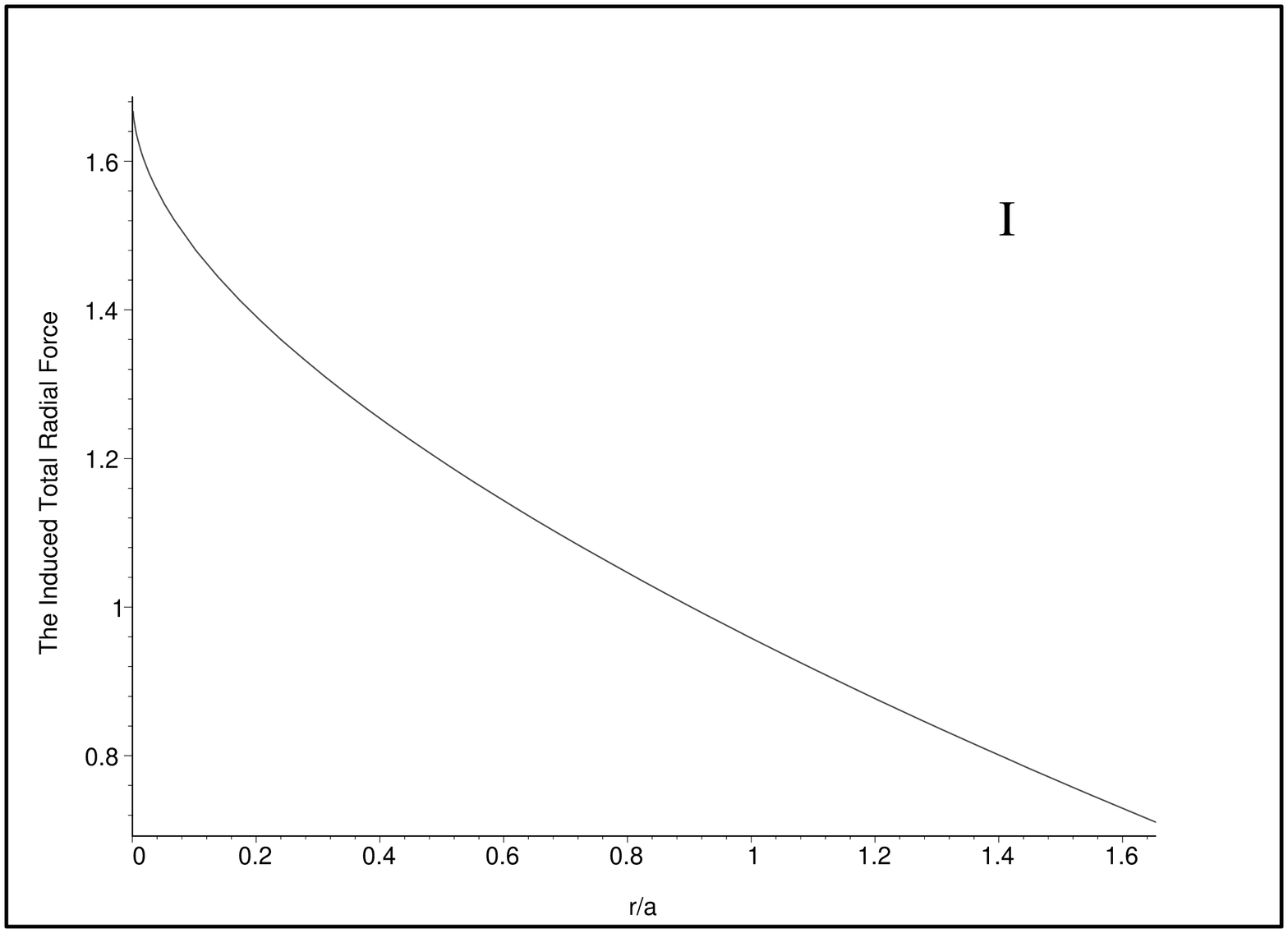}} &
      \resizebox{60mm}{!}{\includegraphics[angle=270]{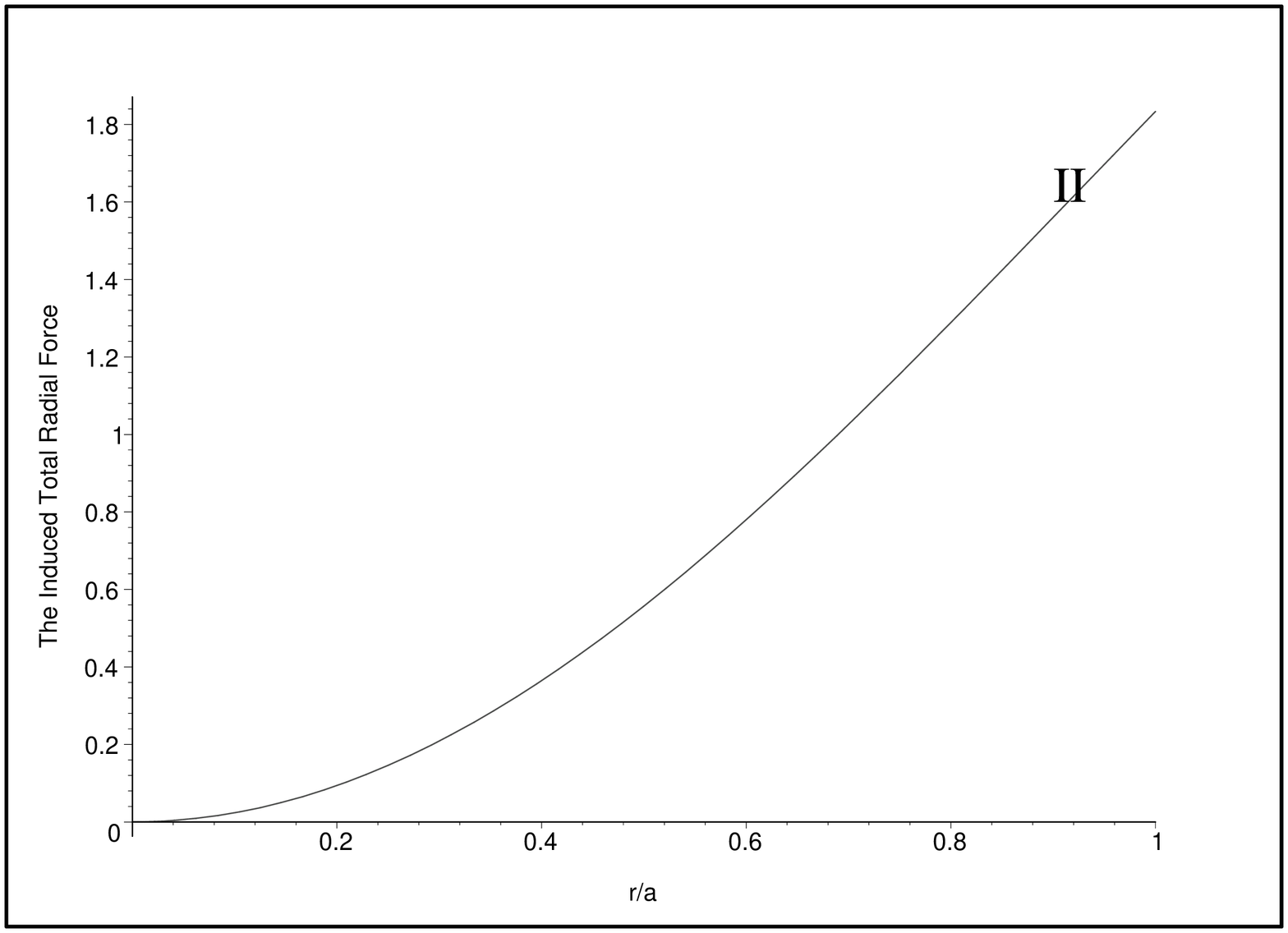}} \\
      \resizebox{59mm}{!}{\includegraphics[angle=270]{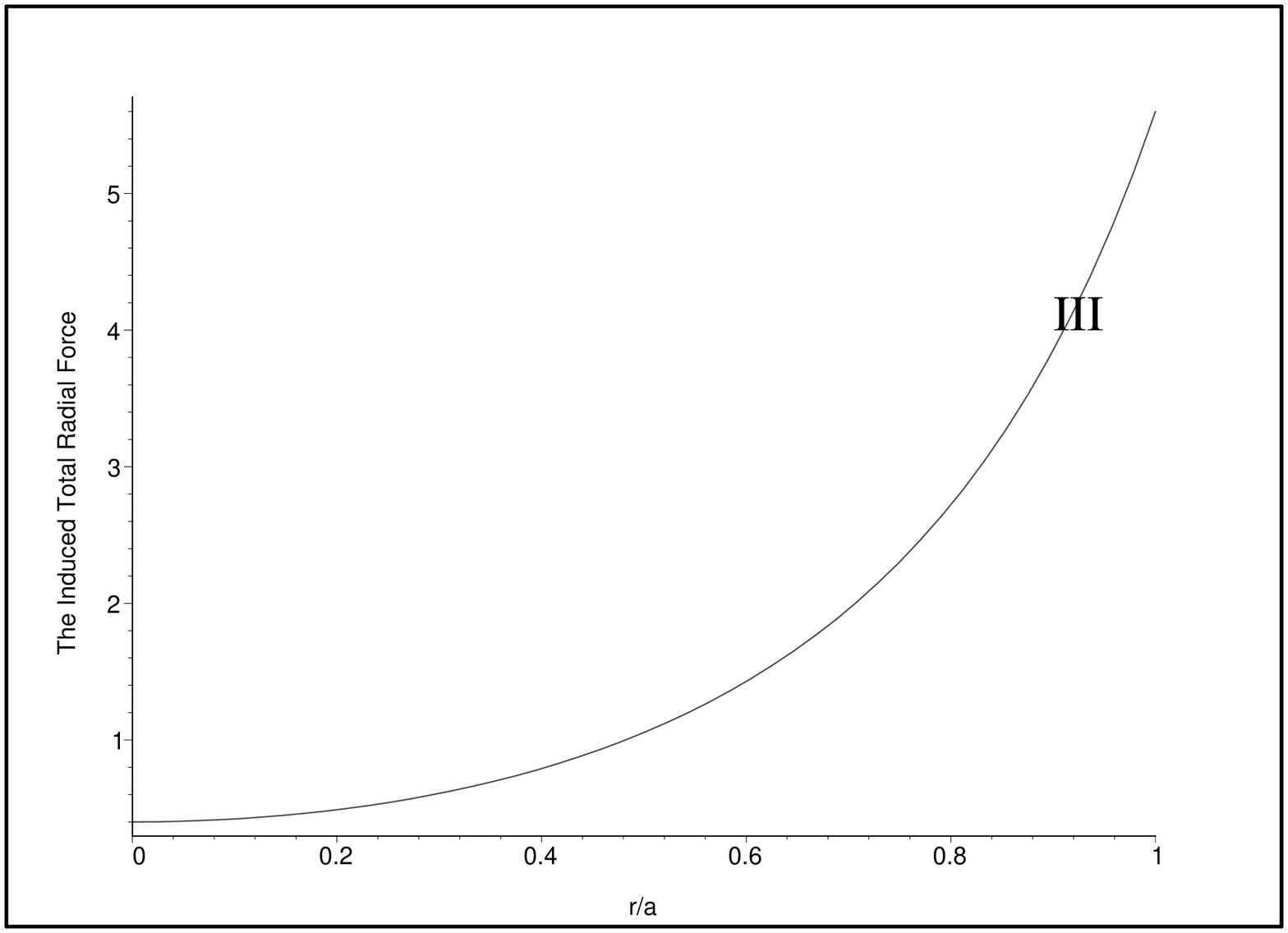}} &
      \resizebox{60mm}{!}{\includegraphics[angle=270]{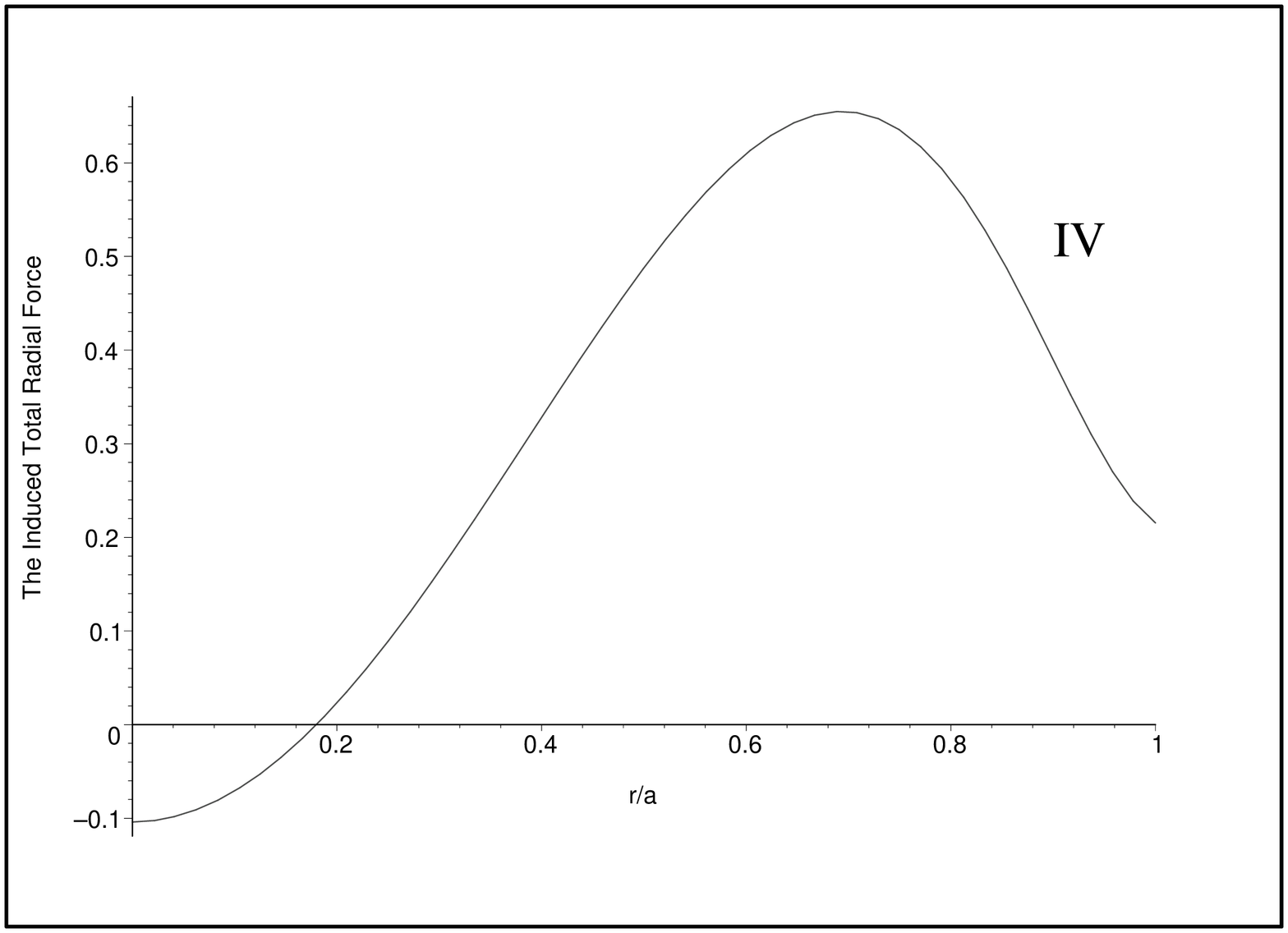}} \\
    \end{tabular}
    \hspace{2pc}
    \begin{minipage}[b]{35pc}
        \caption{The induced total radial force $\tilde{R}$ for the anisotropic configurations. Plots $I$, $II$, $III$ and IV represent Tolman VI, NL Stewart 1, NL Stewart 2 and Gokhroo \& Mehra, respectively. }
        \label{CrackingProfiles}
    \end{minipage}
  \end{center}
\end{figure}

The profiles of $\tilde{R}$ for each model are ploted in Figure \ref{CrackingProfiles}. and the above stability assumptions can be contrasted with the change in sign for the expressions  \ref{FractTolmanVI},  \ref{FractNLStewart1}, \ref{FractNLStewart2} and  \ref{FractGokhroo}.  It is clear from the $\tilde{R}$-plots displayed in this figure, that the models of Tolman VI, NL Stewart 1 and NL Stewart 2 models do not present any cracking point (plates $I$, $II$ and $III$, respectively). On the other hand,  the Florides-Stewart-Gokhroo-Mehra model displays a cracking point at $\eta \approx 0.17986$ within the first potentially unstable region ,$ 0 \lesssim \eta = r/a \lesssim 0.2570$.

\section{Results and conclusions}
\label{Results}
We have revisited the concept of cracking for selfgravitating anisotropic matter configurations introduced by L. Herrera and collaborators \cite{Herrera1992,DiPriscoEtal1994,DiPriscoHerreraVarela1997}. It has been shown that for some particular dependent perturbations, the ratio for fluctuations in anisotropy to energy density, $ {\delta \Delta}/{\delta \rho} $  can be interpreted in terms  of the difference of sound speeds, i.e. $\delta \Delta/\delta \rho~\sim~v^{2}_{s \perp}~-~v^{2}_{s r}$; 
where $v^{2}_{s r}$  and $v^{2}_{s \perp}$ represent the radial and tangential sound speeds, respectively. 
It is evident from \ref{VelocityStability1} that regions where $ v^{2}_{s r}  > v^{2}_{s \perp}$ will be potentially unstable. On the other hand, if $ v^{2}_{s r}  \leq v^{2}_{s \perp}$ everywhere within a matter distribution, no cracking will occur and it could be considered as stable.

This reinterpretation could be useful to refine and make the concept of cracking more physically related to the potential instability due to the behavior of some physical variables within matter configurations. It is easy to determine each sound speed, their difference \ref{SoundRelation} and the sign of the difference. Thereafter, we could clearly identify from \ref{fractura} which regions are more likely to be potentially unstable within a particular matter distribution. This can be appreciated from 
 the Florides-Stewart-Gokhroo-Mehra model (Figure \ref{CrackingProfiles} plate $IV$) which displays a cracking point at $\eta \approx 0.17986$ within the first potentially unstable region $ 0 \lesssim \eta = r/a \lesssim 0.2570$.

 Additionally, because each sound speed has to be less than the speed of light, it implies that their difference has the physical restriction: $| \delta \Delta/\delta \rho|~\sim~|v^{2}_{s \perp}~-~v^{2}_{s r}| \leq 1$. This is very important in order to characterize a particular model as  potentially unstable. It is possible to find cracking points within a configuration for unphysical set of fluctuations in  anisotropy and energy density, i.e. $| \delta \Delta/\delta \rho|~>~1$, but the existence of these cracking points could not lead to physical potentially unstable models. More over, the physical restriction, $| \delta \Delta/\delta \rho|~\leq~1$ also conditions the relative order of magnitude of the perturbations.  

The ratio of perturbations, $\delta \Delta/\delta \rho$ are now not necessarily constant. Models considered in previous works \cite{Herrera1992,DiPriscoEtal1994,DiPriscoHerreraVarela1997,AbreuHernandezNunez2007} have constant fluctuations, because there were no other criteria to introduce in order to evaluate the change in the sign of $\tilde{R}$. Now, the possibility to introduce variable fluctuations based on difference of sound speeds, enrich the applicability of the cracking framework to evaluate instabilities within anisotropic matter configurations. 

 It is worth mentioning that, concerning this criterion, one of the extreme matter configurations mentioned above $P_{\perp}~\neq~0$ and $P_{r}~=~0$ is always potentially stable, and the other $P_{\perp}~=~0$ and $P_{r}~\neq~0$ could experiment a cracking (or overturning) scenario. The study of matter configurations with vanishing radial stresses traces back to G. Lema\^{i}tre \cite{Lemaitre1933} and for non static models have been considered in  \cite{HerreraSantos1997}. Recently, this model has been studied  
\cite{JhinganMagli2000,BarveSinghWitten2000} concerning its relation with naked singularities and conformally
flat models has been considered in \cite{HerreraEtal2001}. Extreme models with vanishing tangential
stresses seems to be useful describing highly compact astrophysical objects having very large magnetic fields ($B\gtrsim10^{15}$ G) \cite{ChaichianEtal2000}

As we have pointed out, any model for a static compact object is worthless if it is unstable against fluctuations of its physical variables. If a particular static model is unstable against these fluctuations it could follow different possible patterns in its subsequent evolution. It could collapse, expand, split or overturn. Perturbations play a crucial role not only evaluating the stability of a particular static model, but identifying trends in possible future evolution of the model. Their study should be considered from different points of view and formalisms. In this work we have considered only those perturbations, related through radial and tangential sound speeds, that lead to identify potentially unstable regions. Independent perturbations (not related via any physical quantity) could also exist and could also lead to cracking (or overturning) points but, in this case there is no criteria to quantify their order of magnitude. Other types of perturbations leading to expanding or collapsing evolutions could be considered in the standard Chandrasekhar's variational formalism ( see \cite{Chandrasekhar1984,DevGleiser2003} and references therein). 
Again, we stress the fact that those different possible evolution patterns for unstable configurations, refers only to a tendency. Its occurrence has to be established from the integration of the full set of Einstein equations. 

\section*{Acknowledgements}
We are indebted to Luis Herrera Cometta for very fruitful discussions concerning the idea of cracking. We gratefully acknowledge the partial financial support of the Consejo de Desarrollo Científico Humanístico y Tecnológico de la Universidad de Los Andes (CDCHT-ULA) under project C-1009-00-05-A, and of the Fondo Nacional de Investigaciones Científicas y Tecnológicas (FONACIT) under projects S1-2000000820 and F-2002000426.
\vspace{2pc}

\bibliographystyle{unsrt}
\bibliography{BiblioLN}

\end{document}